\documentclass[prl,aps,floatfix,twocolumn,showpacs]{revtex4-2} 
\usepackage{color,soul}
\usepackage{graphicx}
\usepackage{amsmath, amsthm, amssymb,mathtools}
\usepackage{amsthm}
\usepackage{multirow}
\usepackage[normalem]{ulem}
\usepackage{soul}
\usepackage{hyperref}
\newcommand{\be}{\begin{equation}}
\newcommand{\ee}{\end{equation}}
\newcommand{\bea}{\begin{eqnarray}}
\newcommand{\eea}{\end{eqnarray}}
\makeatletter

\newcommand{\Rmnum}[1]{\expandafter\@slowromancap\romannumeral #1@}
\makeatother\usepackage{array, makecell} 

\begin{document}

\title{Mpemba effect for a Brownian particle trapped in a single well potential}

\author{Apurba Biswas}
\email{apurbab@imsc.res.in}
\affiliation{The Institute of Mathematical Sciences, C.I.T. Campus, Taramani, Chennai 600113, India}
\affiliation{Homi Bhabha National Institute, Training School Complex, Anushakti Nagar, Mumbai 400094, India}
\author{R. Rajesh} 
\email{rrajesh@imsc.res.in}
\affiliation{The Institute of Mathematical Sciences, C.I.T. Campus, Taramani, Chennai 600113, India}
\affiliation{Homi Bhabha National Institute, Training School Complex, Anushakti Nagar, Mumbai 400094, India}

\date{\today}

\begin{abstract}

Mpemba effect refers to the counterintuitive phenomenon of a hotter system equilibrating faster than a colder system when both are quenched to the same low temperature. For a Brownian particle trapped in a piece-wise linear single well potential that is devoid of any other metastable minima we show the existence of Mpemba effect for a wide range of parameters through an exact solution. This result challenges the prevalent explanation of the Mpemba effect that requires the energy landscape to be rugged with multiple minima. We also demonstrate the existence of inverse and strong Mpemba effects. 
\end{abstract}

\maketitle

\noindent

Mpemba effect describes an anomalous relaxation phenomenon wherein a system that is initially hotter equilibrates faster than a system that is initially cooler, when both systems are quenched to the same low temperature~\cite{Mpemba_1969}. Recently,  there  has been considerable experimental and theoretical interest in the Mpemba effect.  First observed in water~\cite{Mpemba_1969,Mirabedin-evporation-2017, vynnycky-convection:2015, katz2009hot, david-super-cooling-1995, zhang-hydrbond1-2014,tao-hydrogen-2017,Molecular_Dynamics_jin2015mechanisms, gijon2019paths}, it has now been experimentally observed in a wide range of physical systems such as magnetic alloys~\cite{chaddah2010overtaking}, polylactides~\cite{Polylactide}, clathrate hydrates~\cite{paper:hydrates},  colloidal systems~\cite{kumar2020exponentially,kumar2021anomalous,bechhoefer2021fresh}, etc. Theoretical studies, focusing on model systems, have shown the existence of Mpemba effect in spin systems~\cite{PhysRevLett.124.060602,Klich-2019,klich2018solution,das2021should,PhysRevE.104.044114,teza2021relaxation},  Markovian systems with few states~\cite{Lu-raz:2017,PhysRevResearch.3.043160}, particles diffusing in a potential~\cite{Walker_2021,Busiello_2021,lapolla2020faster,walker2022mpemba,degunther2022anomalous}, active systems~\cite{schwarzendahl2021anomalous}, spin glasses~\cite{SpinGlassMpemba}, molecular gases in contact with a thermal reservoir~\cite{moleculargas,gonzalez2020mpemba,gonzalez2020anomalous,PhysRevE.104.064127}, quantum systems~\cite{PhysRevLett.127.060401,chatterjee2023quantum,nava2019lindblad}, systems with phase transitions~\cite{holtzman2022landau,das2021should,zhang2022theoretical,teza2022eigenvalue}, and granular systems~\cite{Lasanta-mpemba-1-2017,Torrente-rough-2019,mompo2020memory,PhysRevE.102.012906,biswas2021mpemba,biswas2022mpemba,megias2022mpemba,biswas2023measure}.

Several system-specific reasons have been proposed to explain the Mpemba effect. For example, the different reasons proposed for Mpemba effect in water include evaporation~\cite{Mirabedin-evporation-2017}, convection~\cite{vynnycky-convection:2015}, dissolved gases~\cite{katz2009hot}, supercooling~\cite{david-super-cooling-1995}, hydrogen bonding~\cite{zhang-hydrbond1-2014,tao-hydrogen-2017,Molecular_Dynamics_jin2015mechanisms} and non-equipartition of energy~\cite{gijon2019paths}. For clathrate hydrates~\cite{paper:hydrates} interplay between evaporation and properties of the hydrogen bonds has been suggested while Mpemba effect in  magnetic alloys~\cite{chaddah2010overtaking} has been attributed to the kinetic arrest of non-equilibrium phase during the relaxation. However, a general understanding of the origin of Mpemba effect is lacking. Recently, insights obtained from the theoretical studies, in particular the analytically tractable models with only few degrees of freedom, suggest that the main driver of the Mpemba effect is the ruggedness or the presence of multiple minima in the energy landscape. In particular, it has been suggested that a metastable minimum, in addition to the global minimum, tends to trap a system at lower temperature more effectively than a system at higher temperature, resulting in a faster relaxation of the hotter system~\cite{Lu-raz:2017,PhysRevLett.124.060602,Klich-2019,Walker_2021,schwarzendahl2021anomalous,biswas2023mpemba}. This viewpoint was further supported  in a recent experiment on a single Brownian particle diffusing in an asymmetric  double well harmonic potential with linear slopes near the boundaries of the domain, where Mpemba effect was clearly  demonstrated~\cite{kumar2020exponentially}. However, the necessity of a metastable minimum for the Mpemba effect has been questioned  in recent works~\cite{Walker_2021,biswas2023mpemba}. For a particular choice of a piece-wise constant potential, it was shown that the Mpemba effect is observed when the metastable state has neutral equilibrium~\cite{Walker_2021}, while its existence was shown for the particular case of  a double well potential when the metastability is just lifted~\cite{biswas2023mpemba}.

In this paper, we solve exactly for the relaxation dynamics of a Brownian particle in a piece-wise linear single well potential. By obtaining the phase diagram for different combinations of the parameters defining the potential, we show that the Mpemba effect is observable for a wide choice of potentials, conclusively showing that  the origin of the Mpemba effect  does not require the energy landscape to have metastable minima, in addition to the global minimum. We also demonstrate the existence of inverse Mpemba effect for systems that are heated, as well as the strong Mpemba effect when the colder system cools exponentially faster. In addition, we show numerically the presence of the Mpemba effect for an overdamped particle in a single well asymmetric harmonic potential that can be realized in the experimental set up similar to that of Ref.~\cite{kumar2020exponentially}.

We consider a Brownian particle in one dimension trapped in a single well potential  $\tilde{U}(\tilde{x})$ that is infinite outside a domain of length $L$. The thermal environment  is characterized by noise $\eta$ that has the characteristics $\langle \eta(\tilde{t}) \rangle=0$ and $\langle \eta(\tilde{t}) \eta(\tilde{t}') \rangle= 2 \gamma k_B \tilde{T}_b \delta (\tilde{t}-\tilde{t}')$, where $\gamma$ is the damping coefficient,  $\tilde{t}$ is time,  $\tilde{T}_b$ is the temperature of the thermal bath and $k_B$ is the Boltzmann's constant. The motion of the particle is described by the  Langevin equation
\be
\gamma \frac{d\tilde{x}}{d\tilde{t}}=-\frac{d\tilde{U}}{d\tilde{x}}+\eta(\tilde{t}). \label{langevin eqn}
\ee

We consider dimensionless variables $x=(2\pi/L) \tilde{x}$, $T=\tilde{T}/\tilde{T}_b$, $U=\tilde{U}/(k_B \tilde{T}_b)$ and $t=(4 \pi^2 k_B \tilde{T}_b/\gamma L^2)\tilde{t}$. 
The corresponding Fokker-Planck (FP) equation for the evolution of the probability density $p(x,t)$ in the non-dimensionalised variables  is given by~\cite{risken1996fokker,morsch1979one}
\be
\frac{\partial p(x,t)}{\partial t}=\frac{\partial}{\partial x}\Big[\frac{d U(x)}{dx} p(x,t) \Big]+\frac{\partial^2 p(x,t)}{\partial x^2}=\mathcal{L}_{FP}p(x,t), \label{FP eqn1}
 \ee
where $\mathcal{L}_{FP}$ is the FP operator: 
\be
\mathcal{L}_{FP}=\partial_x U' +\partial^2_x.
\ee 

\begin{figure}
\centering
\includegraphics[width= 0.7\columnwidth]{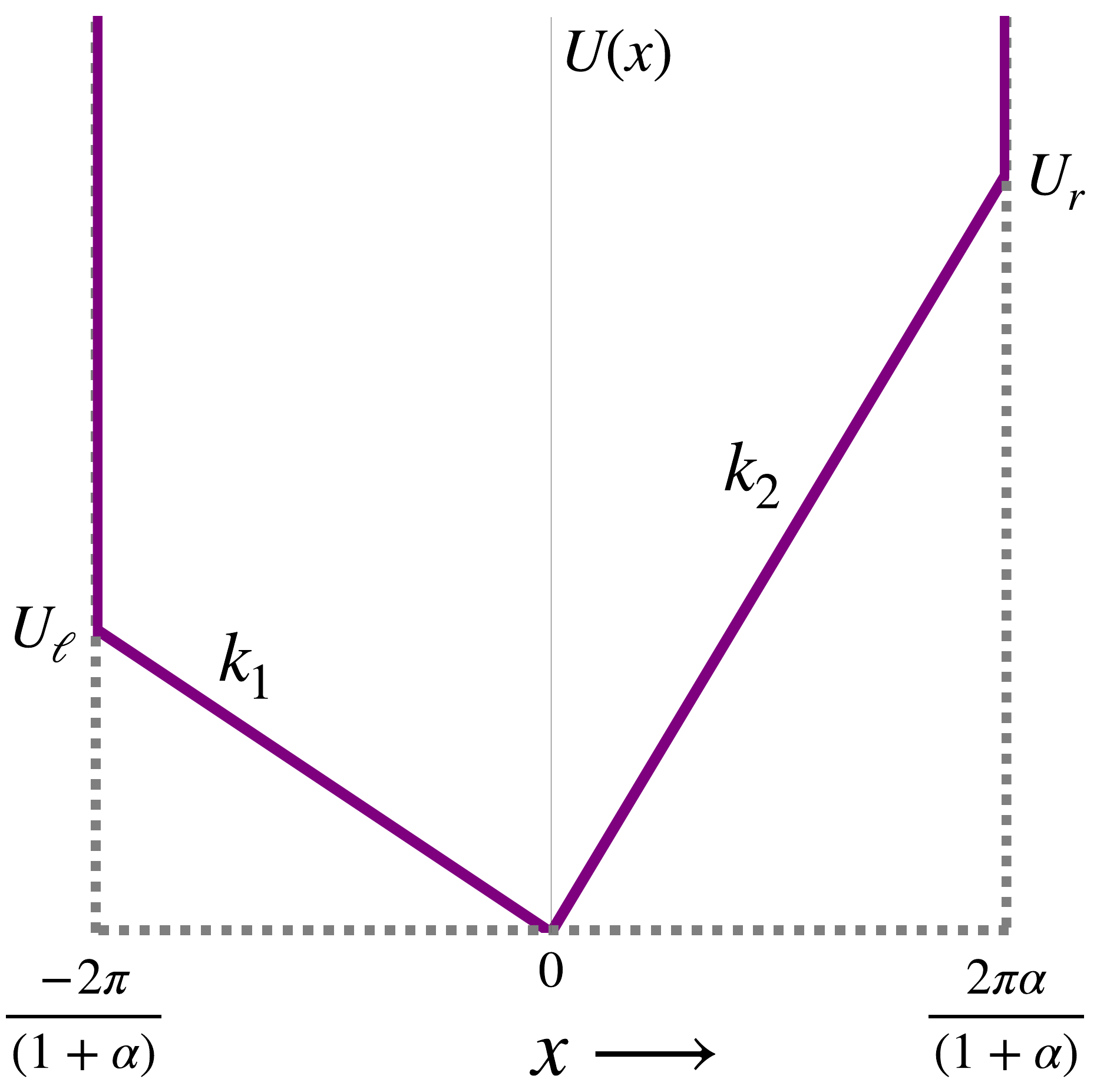} 
\caption{\label{potential shape}Schematic diagram of the piecewise linear single well potential. The parameters $k_1$ and $k_2$ refer to the slopes, $\alpha$ denotes the ratio of width of right domain to the left domain,  and $U_{\ell}$ and $U_r$ are the values of the potential at the boundaries. }
\end{figure}

With the aim  to demonstrate and characterize  Mpemba effect in the absence of metastable states, we consider a single well potential which is piece-wise linear, as shown in Fig.~\ref{potential shape}, and is given by
\be
U(x)=
\begin{cases}
U_{\ell}+k_1 (x+\frac{2\pi}{1+\alpha}), & \frac{- 2\pi}{1+\alpha} < x < 0 \\
k_2 x , & 0 < x < \frac{2 \pi \alpha}{1+\alpha},  \label{potential form}
\end{cases}
\ee
where $k_1$, $k_2$ are the slopes and  $U_{\ell}$, $U_r$ are the values of the potential at the boundaries.  The minimum of the potential, set equal to zero, is fixed at $x=0$. The parameter $\alpha$ is the ratio of the length of the right domain to that to the left domain. The piece-wise linearity makes the problem analytically tractable.

To solve the FP equation in Eq.~(\ref{FP eqn1}), we closely follow the known methods in Refs.~\cite{risken1996fokker,morsch1979one} and the solution in Ref.~\cite{biswas2023mpemba}. We first transform the  associated FP operator $ \mathcal{L}_{FP}$ to an self-adjoint operator $\mathcal{L}$:
 \be
 \mathcal{L}=e^{U(x)/2} \mathcal{L}_{FP} e^{-U(x)/2}=\frac{\partial^2}{\partial x^2}-\frac{1}{4}\Big(\frac{dU}{dx}\Big)^2+ \frac{1}{2}\frac{d^2 U}{dx^2}. \label{adjoint FP operator}
 \ee
The problem then reduces to solving the eigenvalue equation
\be
\mathcal{L} \psi_n =-|\lambda_n| \psi_n, \label{eigenvalue eqn}
\ee
where $ \psi_n$ are the eigenfunctions of the  operator $\mathcal{L}$ corresponding to the eigenvalue $\lambda_n$. Note that both the operators $\mathcal{L}$ and $\mathcal{L}_{FP}$ have the same eigenvalue $\lambda_n$ but their respective eigenfunctions $\psi_n(x)$ and $\phi_n(x)$ are related by $\psi_n(x)=\exp\left(U(x)/2\right) \phi_n(x)$. The details of the solution of the Eq.~(\ref{eigenvalue eqn}) for the eigenvalues and eigenfunctions  can be found in Supplemental Information~\cite{supp}.

Given the eigenspectrum, the probability distribution function $p(x,t)$ is given by 
\be
p(x,t)=\frac{e^{-U(x)}}{\mathcal{Z}(1)}+\sum_{n\geq 2} a_n e^{\frac{-U(x)}{2}} \psi_n(x) e^{-|\lambda_n| t}, \label{prob density solution}
\ee
where the coefficients are fixed by the initial distribution $p(x',0)$: $a_n=\int dx' ~p(x',0) ~e^{\frac{U(x')}{2}} ~\psi^{*}_n(x')$. 
The first term in the right hand side of Eq.~(\ref{prob density solution}) corresponds to the eigenvalue $\lambda_1=0$ and describes the final equilibrium Boltzmann distribution with partition function $\mathcal{Z}(1)=\int e^{-U(x)} dx$ at the bath temperature $T_b=1$. Since the eigenvalues $\lambda_n$ of the FP operator follow the order: $\lambda_1=0>\lambda_2>\lambda_3 \ldots$, we can approximate for $p(x,t)$ at large times as
\be
p(x,t)\simeq \frac{e^{-U(x)}}{\mathcal{Z}(1)}+  a_2 e^{\frac{-U(x)}{2}} \psi_2(x) e^{-|\lambda_2| t},~t \gg \frac{1}{|\lambda_3|}. \label{prob density approx}
\ee
\begin{figure}
\centering
\includegraphics[width=\columnwidth]{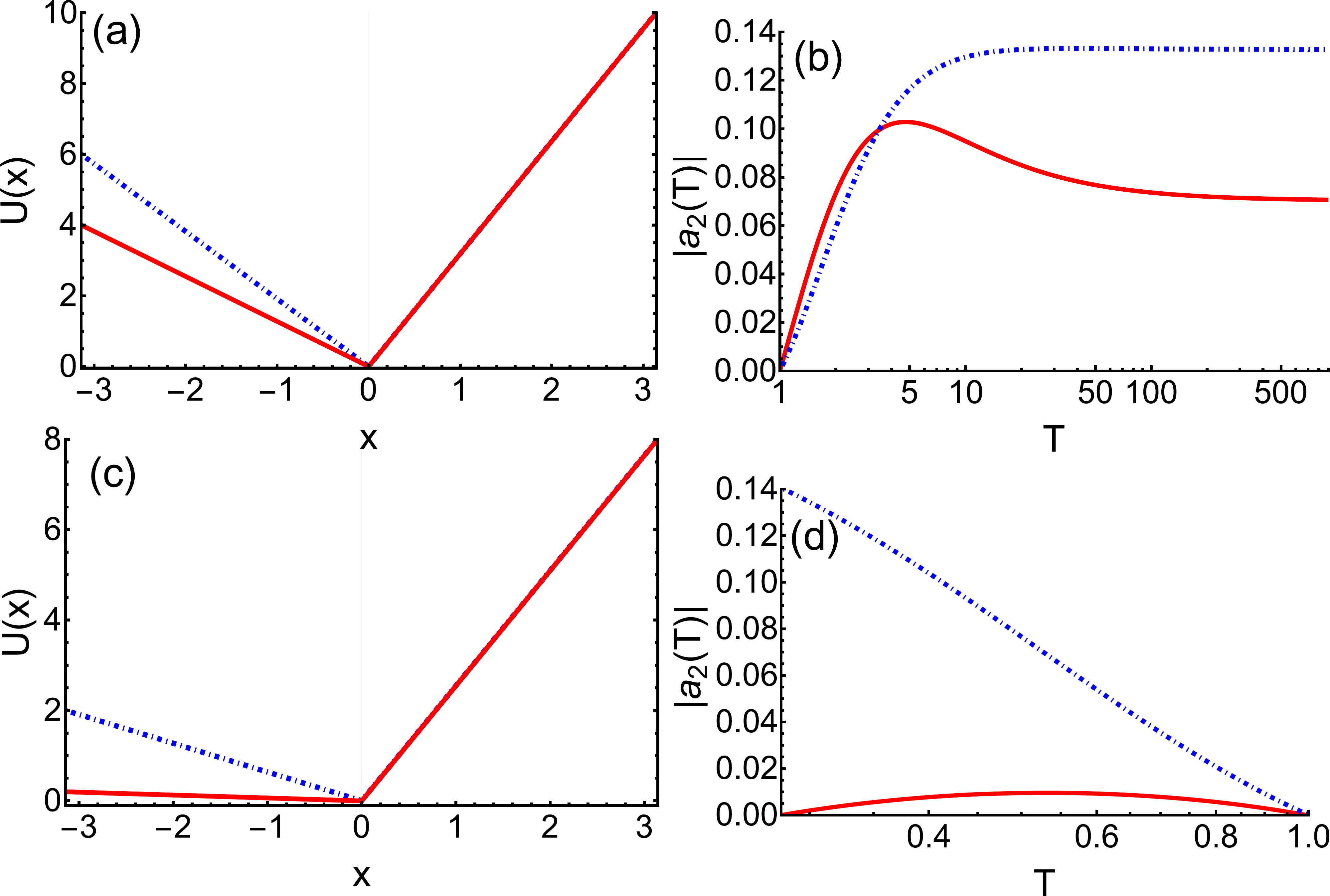}
\caption{Illustration of the Mpemba effect and the inverse Mpemba effect for particular choices of the potential $U(x)$.   (a) Two choices of the potential corresponding to  $U_{\ell}=4.0$, (red) and $U_{\ell}=6.0$, (blue) keeping $U_r=10.0$, $\alpha=1.0$ fixed. (b) For $U(x)$ in (a) with smaller (larger) value of $U_{\ell}$,  $|a_2(T)|$  is non-monotonic (monotonic) for  $T>1$, demonstrating presence (absence)  of the Mpemba effect. (c) Two choices of the potential corresponding to  $U_{\ell}=0.2$, (red) and $U_{\ell}=2.0$ (blue) keeping  $U_r=8.0$, $\alpha=1.0$ fixed.
(d) For $U(x)$ in (c) with smaller (larger) value of $U_{\ell}$,  $|a_2(T)|$  is non-monotonic (monotonic) for  $T<1$, demonstrating presence (absence)  of the inverse Mpemba effect.}\label{fig combined symmetric}
\end{figure}
\begin{figure*}
\includegraphics[width=\textwidth]{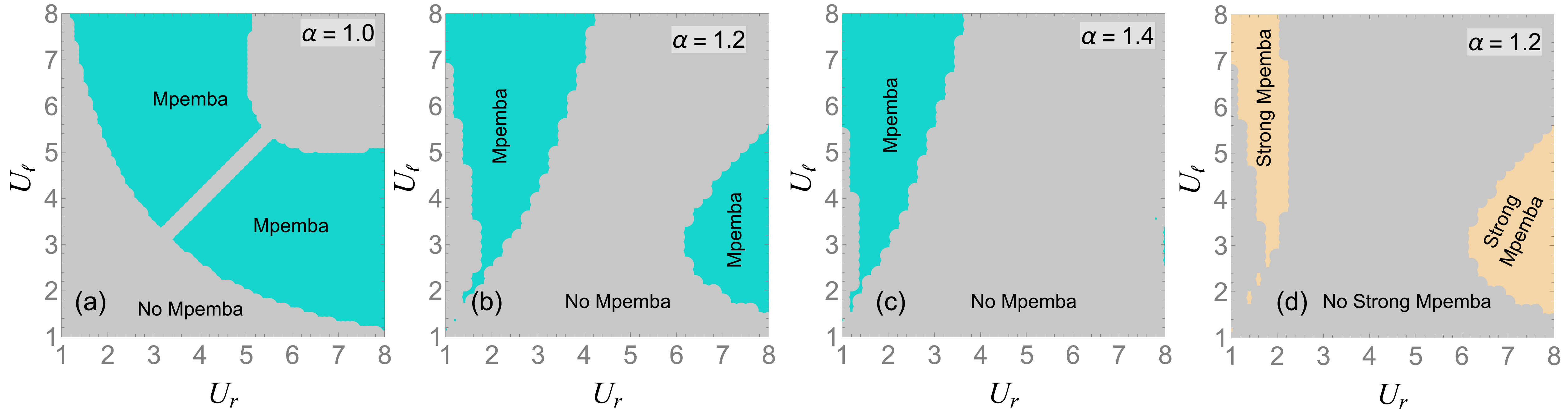} 
\caption{\label{phase diagram asymmetric}The $U_\ell$--$U_{r}$ phase diagram showing regions where Mpemba effect is present (green) and absent (grey) for (a) $\alpha=1.0$, (b) $\alpha=1.2$, and (c) $\alpha=1.4$. (d) The $U_\ell$--$U_{r}$ phase diagram showing regions where strong Mpemba effect is present (yellow) and absent (grey) for  $\alpha=1.2$. }
\end{figure*}

We will now use Eq.~(\ref{prob density approx}), that describes the relaxation dynamics at late times, to quantitatively define Mpemba effect for this system. Consider two systems, $P$ and $Q$ initially in equilibrium at temperatures $T_h$ and $T_c$ respectively, where  $T_h > T_c$.  Let $\pi(T)$ denote the corresponding equilibrium probability distribution at temperature $T$. Both the systems are then simultaneously quenched to a common bath temperature $T_b=1$ where $T_h > T_c>1$. If $P$ equilibrates faster than $Q$, the Mpemba effect is said to exist. Faster equilibration is quantified in terms of a distance from equilibrium function $D[p(t),\pi(1)]$ which measures the distance between the instantaneous distribution $p(x,t)$ from the final equilibrium Boltzmann distribution, $\pi(1)$. It has been argued~\cite{Lu-raz:2017}  that the existence of Mpemba effect is independent of the choice of $D[p(t),\pi(1)]$ provided that the distance measure obeys certain properties. These are (i) If  $T_h>T_c>1$, then the distance function should satisfy $D[\pi(T_h), \pi(1)]>D[\pi(T_c), \pi(1)]$, i.e., higher the temperature larger the distance, (ii) $D[p(t), \pi(1)]$ should be a non-increasing function of time, and (iii) $D[p(t), \pi(1)]$ should be a convex function of $p(t)$.

A convenient choice of such a distance measure is the total variation distance defined as $D[p(t),\pi(1)] \equiv L_1(t)=\int dx |p(x,t)-\pi(x,1)|$. When $t=0$, since $T_h > T_c$, initially $L^h_1>L^c_1$. For Mpemba effect to exist,  we require that $L^h_1<L^c_1$ at late times. In Eq.~(\ref{prob density approx}), since $\lambda_2$ is independent of initial condition,  the above condition for Mpemba effect reduces to   $|a^c_2|>|a^h_2|$~\cite{Lu-raz:2017,Klich-2019}. Equivalently, if $|a_2(T)|$ is non-monotonic with temperature, then we would always be able to make a choice of $T_h$ and $T_c$ such that $|a^c_2|>|a^h_2|$.

The methodology we employ is as follows. Given a potential, we solve for the eigenspectrum,  which in turn allows us to compute the time evolution of the probability distribution of the particle during equilibration  using Eq.~(\ref{prob density solution}). We then determine $|a_2(T)|$ for $T> 1$ and check for non-monotonicity. A similar analysis can be done for the inverse Mpemba effect in which the systems are quenched to a temperature that is higher than the initial temperatures. To check for  presence of inverse Mpemba effect, we check whether $|a_2(T)|$ for $T< 1$ is non-monotonic.

We first show the existence of Mpemba effect and inverse Mpemba effect for a single well potential in order to demonstrate that the origin of the effect does not require more than one energy minima.  For this purpose, we make specific choices of the single well potential, as shown in Fig.~\ref{fig combined symmetric}. For the two instances of $U(x)$, shown in Fig.~\ref{fig combined symmetric}(a), $|a_2(T)|$ for $T>1$ is non-monotonic for one and monotonic for the other, showing the presence and absence of Mpemba effect depending on the choice of potential. Similar construction is possible for the inverse Mpemba effect. For the two instances of $U(x)$, shown in Fig.~\ref{fig combined symmetric}(b), $|a_2(T)|$ for $T<1$ is non-monotonic for one and monotonic for the other, showing the presence and absence of inverse Mpemba effect depending on the choice of potential. Beyond showing its existence, we will not discuss the inverse Mpemba effect further.

To get an insight of what potentials allow for the Mpemba effect, we now construct phase diagrams demarcating regions that show Mpemba effect from regions that do not. The potential $U(x)$ is characterized by three parameters: $U_{\ell}$, $U_r$, and $\alpha$. Asymmetry in the potential can be introduced through $U_{\ell}\neq U_r$ and/or $\alpha \neq 1$. The fully symmetric potential is not expected to show Mpemba effect~\cite{Walker_2021,degunther2022anomalous}. To explore the effect of all three parameters, we determine the phase diagram in the $U_\ell$--$U_r$ plane for fixed $\alpha$. The phase diagrams for $\alpha=1.0$, $1.2$, and $1.4$ are shown in Fig.~\ref{phase diagram asymmetric}(a)--(c). First, we observe that, for all three values of  $\alpha$, the fraction of the parameter space that shows Mpemba effect is not negligible. This implies that, for the single well potential, Mpemba effect can be observed for generic choices of potentials.  Second, we observe that when $U_\ell=U_r$, Mpemba effect does not exist even if asymmetry is introduced through $\alpha \neq 1$.  Thus asymmetry through only different domain widths is not sufficient. Third, when $\alpha$ is increased, thereby increasing asymmetry in the potential, the area of the region showing Mpemba effect decreases. This is contrary to expectation based on the experiment on colloids~\cite{kumar2020exponentially} where it was observed that asymmetry in domain widths in double well potentials enhanced Mpemba effect. To further demonstrate that increasing $\alpha$ may not be beneficial, we analyzed the particular case of $U_\ell=2.0$ and $U_r=8.0$ and find that Mpemba effect is present for $\alpha$ only in the domain that is approximately $(0.3, 1.1)$.

We now discuss the existence of strong Mpemba effect for single well potentials. Strong Mpemba effect refers to the case when the hotter system cools exponentially faster than the colder system. In terms of Eq.~(\ref{prob density approx}), this can be achieved only if the coefficient $a_2(T)$ is zero. When $a_2=0$,  the slowest relaxation mode does not contribute. For $\alpha=1.2$, we identify those potentials for which $a_2$ is zero for some temperature. Then, if we choose $T_c$ to be the same temperature, we would observe strong Mpemba effect. Figure~\ref{phase diagram asymmetric}(d) shows the phase space region in the $U_\ell -U_r$ plane where the strong Mpemba effect is present/absent for $\alpha=1.2$. Thus, even for the existence of strong Mpemba effect, a rugged energy landscape is not necessary.

\begin{figure}
\includegraphics[width=\columnwidth]{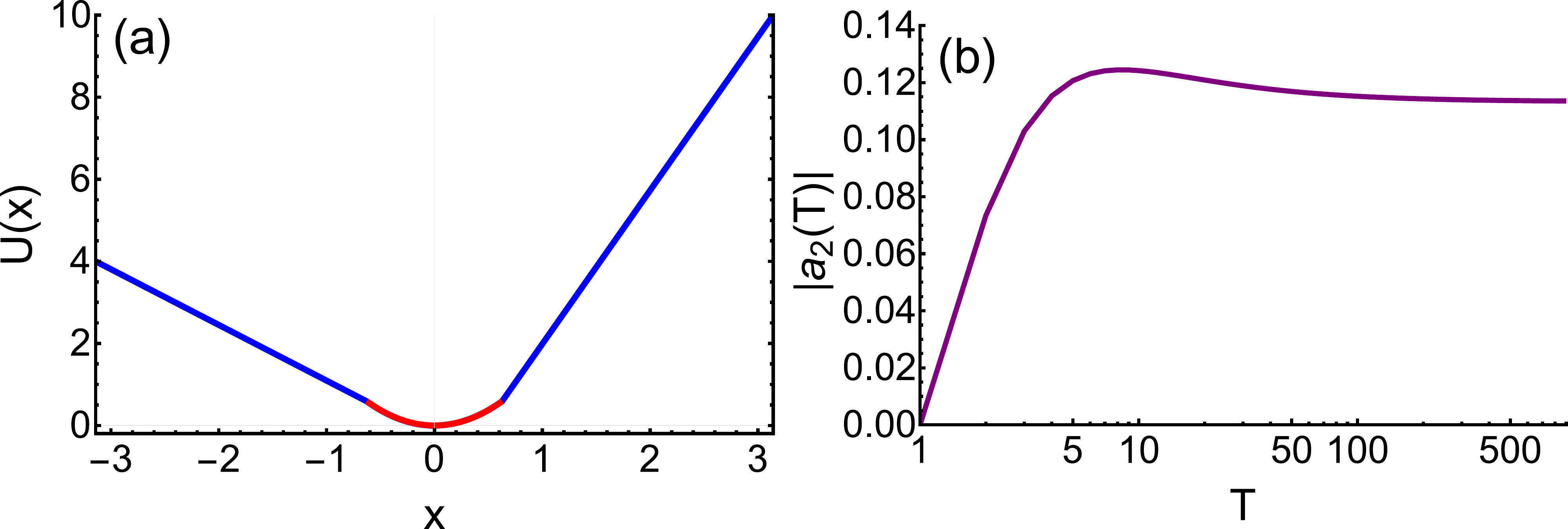}
\caption{Numerical results for the illustration of the Mpemba effect for a  single well harmonic potential.   (a) Shape of the potential with harmonic  well (red region)and linear slopes near the boundaries (blue regions).  (b) The variation of $|a_2(T)|$ with temperature $T$, obtained by a numerical solution, is non-monotonic, showing the presence of the Mpemba effect.} \label{fig harmonic trap}
\end{figure}

Finally, we discuss the connection to experiments.  In the experiment of Kumar and Bechhoefer~\cite{kumar2020exponentially}, Mpemba effect was demonstrated for a colloidal particle trapped in an asymmetric double well harmonic potential with linear behavior near the edges.  We now show that the piece-wise linear single well potential considered in this paper, when modified to have a harmonic minimum and linear behavior near the edges, continues to exhibit Mpemba effect. Consider the particular realization of the  single well potential, shown in Fig.~\ref{fig harmonic trap}(a), which has a harmonic minimum. The precise details of the potential are given in \cite{supp}. However, for such potentials, it is no longer possible to analytically solve for  the eigenspectrum. Instead, we can solve the eigenspectrum numerically, and thus obtain the coefficient $a_2$.   For the particular choice of potential, $|a_2(T)|$ is clearly non-monotonic [see Fig.~\ref{fig harmonic trap}(b)], showing the existence of the Mpemba effect in single well harmonic potentials. Thus, we expect that if the experiment in Ref.~\cite{kumar2020exponentially} is repeated with a single well potential, Mpemba effect will  be observed.

In summary, we solved exactly the relaxation dynamics of a Brownian particle in an asymmetric single well potential that is piecewise linear. By identifying the regions in parameter space that exhibits the Mpemba effect, we not only show that the presence of the Mpemba effect does not require the energy landscape to have multiple minima, but also that the Mpemba effect can be realized for generic choices of potentials.  We also demonstrated that both the inverse and strong Mpemba effects can also be realized for single well potentials. We also showed numerically that single well harmonic potentials, as opposed to piece-wise linear single well potentials, continue to exhibit Mpemba effect, opening up the possibility of experimental realization along the lines of the experiment in Ref~\cite{kumar2020exponentially}.

We thank S. Vemparala for a critical reading of the manuscript.


%

\end{document}